\documentclass[
 reprint,
 amsmath,amssymb,
 aps,
prb,
]{revtex4-2}

\usepackage{graphicx}% Include figure files
\usepackage{dcolumn}% Align table columns on decimal point
\usepackage{bm, xcolor}% bold math
\usepackage[hidelinks]{hyperref}% add hypertext capabilities
\hypersetup{
  colorlinks   = true, %Colours links instead of ugly boxes
  urlcolor     = blue, %Colour for external hyperlinks
  linkcolor    = blue, %Colour of internal links
  citecolor   = blue %Colour of citations
}
\begin{document}

%\preprint{APS/123-QED}

\title{%(i) Universal polarization imprints around acoustic zeros \\
%(ii) Three-dimensional polarization near field nodes of acoustic waves \\
Non-Divergent Spinning Substructures Near Acoustic Field Nodes}

\author{Andrew Kille}
\author{Andrei Afanasev}
 \email{afanas@gwu.edu}
\affiliation{Department of Physics, The George Washington University, Washington, DC 20052, USA
}

\date{\today}

\begin{abstract}
In this work, we examine the extraordinary behavior of polarization and spin angular momentum (AM) density in the vicinity of longitudinal field zeros in three-dimensional monochromatic acoustic fields. We demonstrate that, as governed by the continuity equation, the velocity fields of arbitrary acoustic sources maintain non-diffractive elliptical polarization structures that enclose longitudinal field zeros, despite having divergent transverse spatial profiles of intensity. Furthermore, embedded in these nonparaxial field contours, for infinite distance, are threads of circular polarization singularities. We illuminate these inherent properties in acoustic vortex fields, dipole arrays, and the famous Young's double slit experiment. Our results reveal novel characteristics of vector sound waves that provide a platform for future studies and applications of structured acoustic waves and chiral acoustic phenomena.
\end{abstract}

\maketitle

%\tableofcontents

\section{\label{sec:intro}Introduction}
Singularities in complex inhomogeneous scalar and vector wave fields exhibit rich topological characteristics.  In classical vector waves, there exists a fundamental property known as polarization, in which an abundance of singular features have been studied extensively, particularly those found in structured monochromatic light \cite{berry2023,dennis2009}. Structured acoustic waves with twisted wave fronts carrying orbital angular momentum were first demonstrated experimentally in Ref.~\cite{hefner1999acoustical}. Recent theoretical and experimental results demonstrate analogous properties of polarization and angular momenta in the longitudinal (curl-free) vector velocity fields of acoustic waves \cite{courtney2013dexterous,muelas-hurtado2022, bliokh2019,bliokh2021, shi2019, ge2021}, opening an avenue for future applications in acoustic tweezers \cite{ozcelik2018, baudoin2020,baresch2016, li2021}, acoustofluidics \cite{fan2022, friend2011}, underwater communications \cite{li2020,heidemann2012}, and biomedical imaging \cite{sarvazyan2013, orazbayev2020,rufo2022}. 

In paraxial sound, the velocity field vector is approximately collinear with wave vector $\mathbf{k},$ yielding a homogeneous distribution of linear polarization in space. This constraint is reasonable for distances far from a localized acoustic source, however, in the near-field, all spatial field components are substantial (i.e., nonparaxial in nature). The local rotational field trajectories associated with nonparaxial sound is highly nontrivial \cite{bliokh2019, burns2020}, and can be described by a dynamical variable known as spin angular momentum (AM) density. Peculiar manifestations of spin AM density arise both in both optics and acoustics, notably, the signature of transverse spin, for example, in evanescent waves \cite{eismann2020, bliokh2014, bliokh2019} and two-wave interference \cite{bekshaev2015}.

Threads of polarization singularities, such as lines of strictly circular and linear polarization, are imprinted in vector waves due to the natural occurrence of field zeros. It was shown recently \cite{afanasev2023} that %strongly-focused 
optical vortex beams possess remarkable and non-intuitive features of non-diffractive polarization near the phase singularities for arbitrary large distances. Moreover, a study found that such polarization and momentum structures exist around any persisting transverse field zero (thread of linear polarization) of electromagnetic radiation and light interference patterns from an arbitrary localized source \cite{vernon2023}. Evidently, this general phenomenon of light originates from Maxwell electromagnetism and the stability of electromagnetic nulls in the far-field.

In this text, motivated by these studies \cite{afanasev2023, vernon2023}, we analyze the general properties of polarization and spin near singular imprints of nonparaxial monochromatic acoustic waves. We show that intrinsic field zeros (phase singularities) produce remarkable non-divergent vector characteristics of sound, in direct analogy with features found in light. Indeed, accompanying polarization structures of invariant and stable cross-section formed by the vector velocity fields from a localized acoustic source is persistent lines of polarization singularities. To illustrate these results, we consider nonparaxial, diverging acoustic vortex beams, a wavelength-spaced acoustic dipole array, and Young's double slit experiment. 

While we are aware of the studies demonstrating strong nonlinearity near phase singularities of acoustic fields \cite{richard2020twisting}, our present approach is linear, since we consider effects in low-intensity regions in the far field of the interference patterns.

\section{\label{sec: gen-theory}General Theory}

\subsection{\label{sec: background}Background}
In this section, we briefly review relevant equations of motion and dynamical properties of acoustic waves. In addition, we discuss acoustic polarization singularities and define a polarization parameter important for analyzing spatial distributions of individual velocity field components. 

Let us consider monochromatic acoustic waves of frequency $\omega$ in a homogeneous medium of mass density $\varrho_0$ and compressibility $\beta$:
\begin{equation}{\label{eq: acoustic-fields}}
    \nabla \cdot \mathbf{V} = i\beta\omega P, \qquad \nabla P = i\varrho_0\omega \mathbf{V},
\end{equation}
where $P(\mathbf{r})$ and $\mathbf{V}(\mathbf{r})$ are the complex scalar pressure and vector velocity fields, respectively. The vector velocity field is longitudinal ($\nabla \times \mathbf{V} = \mathbf{0}),$ yet in the nonparaxial regime can be generally written in terms of its transverse and longitudinal components: $\mathbf{V}(\mathbf{r}) = \mathbf{V}_{\perp}(\mathbf{r}) + V_{\|}(\mathbf{r})\mathbf{\hat{e}}_{\|},$ where $\mathbf{\hat{e}}_{\|}$ denotes the unit vector collinear with $\mathbf{k}.$ The degree in which $\mathbf{V}(\mathbf{r})$ is circularly-polarized is given by the normalized spin AM density
\begin{equation}{\label{eq: norm-spin-density}}
\mathbf{S} = \frac{\text{Im}\{\mathbf{V}^*\times \mathbf{V}\}}{|\mathbf{V}|^2}.
\end{equation}
In plane acoustic waves, the spin AM density vanishes, however, inhomogeneous fields produce nonzero spin AM density orthogonal to the polarization ellipse. Moreover, the time-averaged energy density of the acoustic fields is
\begin{equation}{\label{eq: energy-density}}
W = \frac{1}{4}\Bigl(\beta|P|^2 + \varrho_0|\mathbf{V}|^2\Bigr).
\end{equation}
Note that the transverse spatial extent of energy density $W$ naturally diverges with respect to distance from a source, with theoretical exceptions such as Bessel and Airy beams.

In direct analogy to the singular behavior around transverse field zeros of electromagnetic fields, vanishing longitudinal velocity of acoustic waves yields nearby velocity fields that are circularly polarized, i.e., the orientation of the polarization ellipse axes is undefined. Note that lines of circular polarization (commonly referred to as C lines \cite{berry2001,nye1987}) exist where $\mathbf{V}(\mathbf{r})\cdot \mathbf{V}(\mathbf{r}) = 0,$ which is equivalent to the spin condition $|\mathbf{S}(\mathbf{r})| = 1$.

Here, we use a quantity introduced in \cite{afanasev2023, vernon2023}, which we label as the ``LT-alignment" parameter $\chi_{\text{LT}}:$
\begin{equation}{\label{eq: intensity-ratio}}
\chi_{\text{LT}} = \frac{|\mathbf{V}_{\perp}|^2 - |V_{\|}|^2}{|\mathbf{V}|^2}.
\end{equation}
Note that $\chi_{\text{LT}}$ provides a laconic description of the relative longitudinal-to-transverse (LT) field distributions in space, e.g., it follows from Eq.~(\ref{eq: intensity-ratio}) that for $V_{\parallel} = 0,$ we have $\chi_{\text{LT}}=1$; likewise, $\mathbf{V}_{\perp} = \mathbf{0}$ gives $\chi_{\text{LT}} = -1.$ It was shown in \cite{afanasev2023, vernon2023} for optical vortex beams and general electromagnetic radiation that the contour $\chi_{\text{LT}}=0$ draws out a non-diffracting tube enclosing a transverse zero (thread of linear polarization) for arbitrary distance. In Section \ref{sec: formalism}, we demonstrate that a similar effect holds, that is, for any physical source of acoustic waves containing longitudinal field zeroes there exists the %nonparaxial 
contour $|\mathbf{V}_{\perp}|^2 = |V_{\parallel}|^2$ of invariant elliptical cross section that extends to infinity. 

\subsection{\label{sec: formalism}Formalism}

Let us consider far-field acoustic radiation from any localized source, in which the scalar pressure field can be expressed in spherical coordinates $(r,\theta,\varphi)$ in the general form
\begin{equation}{\label{eq: pressure-radiation}}
P(\mathbf{r}) = P_0(\theta, \varphi)\frac{e^{ikr}}{r},
\end{equation}
where $P_0(\theta,\varphi)$ is the far-field directivity factor unique to the source geometry. A straightforward derivation from Eq.~(\ref{eq: acoustic-fields}) gives the arbitrary vector velocity field:
\begin{equation}{\label{sec: velocity-radiation}}
\mathbf{V}(\mathbf{r}) = \begin{pmatrix} V_{r} \\ V_{\theta} \\ V_{\varphi} \end{pmatrix} = \begin{pmatrix} V_{r0}(\theta, \varphi) \\ V_{\theta 0} (\theta, \varphi)/r \\ V_{\varphi 0} (\theta, \varphi)/r \end{pmatrix} \frac{e^{ikr}}{r},
\end{equation}
where the angular factors $V_{r0}, V_{\theta 0},$ and $V_{\varphi 0}$ are
\begin{equation}{\label{eq: velocity-angular}}
\begin{pmatrix} V_{r0}(\theta, \varphi) \\ V_{\theta 0} (\theta, \varphi) \\ V_{\varphi 0} (\theta, \varphi) \end{pmatrix} = (i\varrho_0 \omega)^{-1} \begin{pmatrix} P_0(\theta, \varphi) \\[6pt] \dfrac{\partial P_0(\theta, \varphi)}{\partial \theta} \\[6pt] \dfrac{1}{\sin\theta}\dfrac{\partial P_0(\theta, \varphi)}{\partial \varphi} \end{pmatrix}.
\end{equation}
Importantly, the radial component $V_r$ has $1/r$ dependence, while the transverse components $V_{\theta}$ and $V_{\varphi}$ have $1/r^2$ dependence. Albeit, $V_r$ dominates in the far-field, however when $V_{r0} = 0$ (i.e., $P_0 = 0$), the transverse components are non-neglible, enabling nonparaxial polarization near longitudinal velocity nulls. Here, for arbitrary acoustic radiation, the contour $\chi_{\text{LT}} = 0$ is equivalent to the relation
\begin{equation}{\label{eq: contour}}
|V_{r}(\mathbf{r})|^2 = |\mathbf{V}_{\perp}(\mathbf{r})|^2,
\end{equation}
where $|\mathbf{V}_{\perp}|^2 = \mathbf{V}_{\perp}^* \cdot \mathbf{V}_{\perp} = |V_{\theta}|^2 + |V_{\varphi}|^2.$ Let us examine the behavior of the left-hand side of Eq.~(\ref{eq: contour}), specifically near an angular position $(\theta_0, \varphi_0)$ such that $V_{r0}(\theta_0, \varphi_0) = 0.$ Decomposing $V_r$ into its real and imaginary parts $\mathbf{V}_r = (\text{Re}~V_r, \text{Im}~V_r)^{\text{T}}$, a first-order Taylor expansion in the vicinity of $(\theta_0, \varphi_0)$ gives linear behavior $\mathbf{V}_{r}\simeq \widetilde{\mathbf{V}}_r = \mathbf{J}_{\text{V}} \mathbf{u}$, that is, we can express the approximate longitudinal component $\widetilde{\mathbf{V}}_r$ in terms of its Jacobian:
\begin{align}{\label{eq: longitudinal-jacobian}}
\mathbf{J}_{\text{V}} = \begin{pmatrix} \text{Re}\dfrac{\partial V_{r0}}{\partial \theta} & \text{Re} \dfrac{\partial V_{r0}}{\partial \varphi} \\[10pt] \text{Im} \dfrac{\partial V_{r0}}{\partial \theta} & \text{Im} \dfrac{\partial V_{r0}}{\partial \varphi} \end{pmatrix} \frac{e^{ikr}}{r} = \mathbf{J}_0 \frac{e^{ikr}}{r}.
\end{align}

\begin{figure*}
    \includegraphics[width=7.0in]{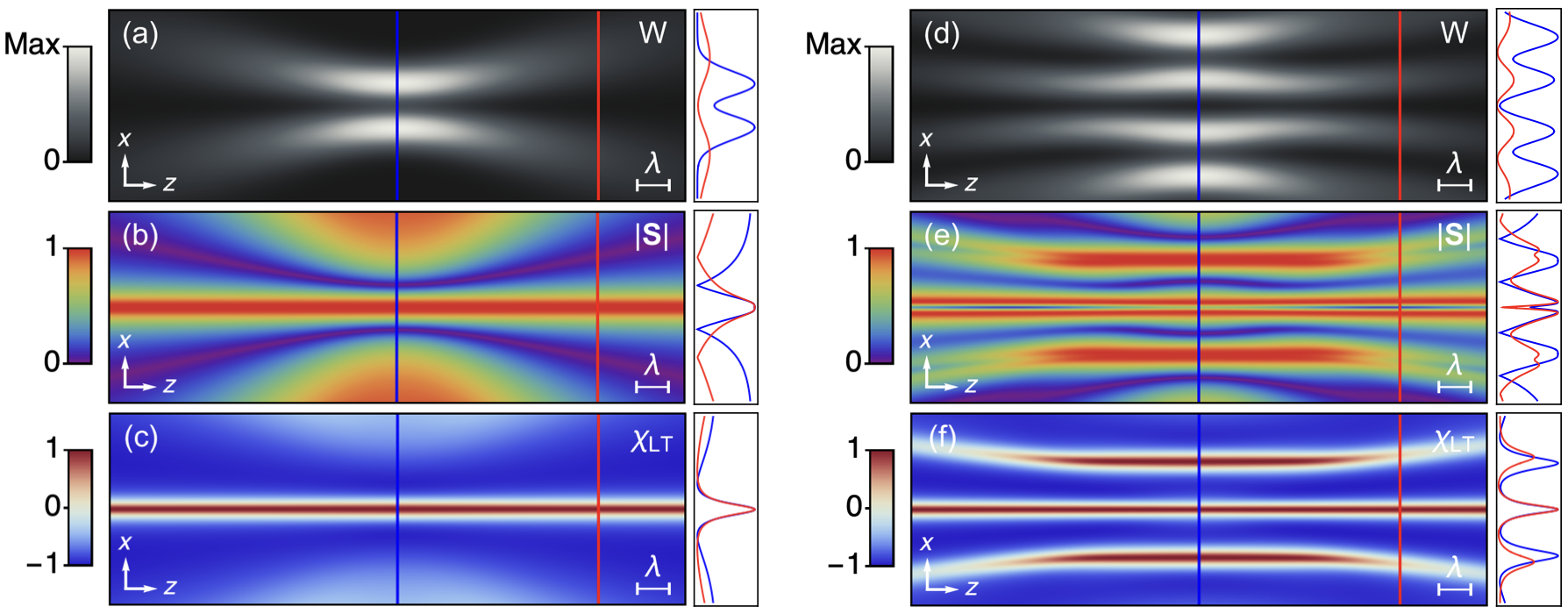}
    \caption{(a)-(c) Energy density $W$, normalized spin AM density $|\mathbf{S}|,$ and LT-alignment parameter $\chi_{\text{LT}}$ for an acoustic Laguerre-Gaussian beam with beam waist $w_0=\lambda$ and orders $\ell=1$, $p=0$ propagating in the $\mathbf{\hat{z}}$-direction. (d)-(f) show the same quantities for non-coaxial superposition of two Laguerre-Gaussian beams ($\ell=1, p=0$), each of beam waist $w_0=\lambda$, with spacing $3\lambda$ along the $\mathbf{\hat{x}}$-axis between their respective centers. In boxes for (a)-(f) are the amplitudes of the corresponding color-density plots at $z=0$ (blue) and $z=2z_R$ (red).}
    \label{fig: LG-plots}
\end{figure*}

Note that $\mathbf{u} = (\theta - \theta_0, \varphi - \varphi_0)^{\text{T}}$ represents the rotated angular space centered around $(\theta_0, \varphi_0).$ The Jacobian $\mathbf{J}_{\text{0}}$ gives a compact, topological description of the dominant longitudinal velocity field near its first-order zeros (i.e., angular positions $(\theta_0, \varphi_0)$ where $\det(\mathbf{J}_{\text{0}}) \neq 0$) \cite{spaegele2023}. Now, Eq.~(\ref{eq: contour}) can be recast in matrix form, dropping radial factors, as
\begin{equation}{\label{eq: matrix-contour}}
\mathbf{u}^{\text{T}}(\mathbf{J}_0^{\text{T}}\mathbf{J}_0)\mathbf{u} = \frac{1}{r^2}|\mathbf{V}_{\perp 0}|^2,
\end{equation}
where $|\mathbf{V}_{\perp 0}|^2 = |V_{\theta 0}|^2 + |V_{\varphi 0}|^2$ is approximately a non-zero constant crossing $(\theta_0, \varphi_0)$, with the condition that $V_r(\theta_0, \varphi_0)$ is a first-order zero. Thus, Eq.~(\ref{eq: matrix-contour}) satisfies the criterion of an ellipse enclosing $(\theta_0, \varphi_0)$, whose cross-section scales directly with $1/r^2,$ thereby exhibiting non-diffractive behavior with respect to radial distance. Geometric properties of the LT-alignment ellipse, which traces out a non-divergent tube for infinite distance, can be derived from Eq.~(\ref{eq: matrix-contour}). Denoting $\lambda_1, \lambda_2$ as the eigenvalues of the matrix $\mathbf{J}_0^{\text{T}} \mathbf{J}_0,$ the semi-axis lengths of the ellipse are
\begin{equation}{\label{eq: semi-axes}}
a = \frac{1}{r}\sqrt{\frac{ |\mathbf{V}_{\perp 0}|^2}{\lambda_1}}, \qquad b = \frac{1}{r}\sqrt{\frac{ |\mathbf{V}_{\perp 0}|^2}{\lambda_2}}.
\end{equation}
It follows from Eq.~(\ref{eq: semi-axes}) that the real-space cross-sectional area of the tube is $A = \pi \sin\theta |\mathbf{V}_{\perp 0}|^2/|\det(\mathbf{J}_0)|,$ which is independent of radial source distance $r.$

In the far field, due to the dominant longitudinal velocity $|\mathbf{V}| \simeq |V_r|$, the C-line condition becomes $\mathbf{V} \cdot \mathbf{V} \simeq V_r^2 = 0.$ This condition is equivalent to intersection of the surfaces $\text{Re}\{\mathbf{V} \cdot \mathbf{V}\} = 0$ and $\text{Im}\{\mathbf{V} \cdot \mathbf{V}\} = 0$, which for brevity we label as $A$ and $B$, respectively. Thus, in the location of $(\theta_0, \varphi_0),$ surfaces $A$ and $B$ are close to each other, however they do not intersect at $(\theta_0, \varphi_0)$ due to the presence of the small transverse field component $\mathbf{V}_{\perp}$. Since $\mathbf{V} \cdot \mathbf{V}$ is approximately quadratic in the far-field, then $A$ and $B$ take the form of either a hyperbola or an ellipse. Thus, with exceptions of additional degrees of symmetry (see Sec. \ref{sec: LG-Beams} for an example in vortex beams), $A$ and $B$ intersect either $2$ or $4$ times in the vicinity of ($\theta_0, \varphi_0$). With increasing radial distance, this approximation becomes nearly exact, so a persistent longitudinal field zero will necessarily have nearby parallel C lines in the far-field.

We summarize this formalism as follows. Despite the inevitable divergence of acoustic fields in space, near a far-field node with angular position $(\theta_0, \varphi_0)$ of any physically realizable source of acoustic radiation, there exists: (i) the nonparaxial contour $|\mathbf{V}_{\perp}| = |V_r|$ that encloses $(\theta_0, \varphi_0)$ in the form of a non-divergent elliptical tube with respect to radial distance, and (ii) accompanying paired threads of circular polarization singularities.

\section{\label{sec: LG-Beams} Acoustic Laguerre-Gaussian Beams}

First, consider nonparaxial vortex beams, which are identifiable by their phase dislocation along the propagation axis and subsequent degree of orbital AM. We construct an acoustic vortex beam by first describing the pressure field with the scalar Laguerre-Gaussian equation in cylindrical coordinates $\mathbf{r} = (\rho, \varphi, z):$
\begin{widetext}
\begin{equation}{\label{eq: LG-pressure}}
P(\mathbf{r}) = A\frac{w_0}{w(z)} \Bigl(\frac{\rho}{w(z)}\Bigr)^{|\ell|} \mathcal{L}^{\ell}_p \exp\Bigl(-\frac{\rho^2}{w^2(z)}\Bigr) \exp\Bigl\{i\Bigl[\ell\varphi + kz + k\frac{\rho^2}{2R(z)} - (|\ell|+2p+1)\xi(z)\Bigr]\Bigr\},
\end{equation}
\end{widetext}
where $A$ is the normalization constant, $w_0$ is the beam waist at $z=0,$ $w(z) = w_0\sqrt{1+(z/z_R)^2}$ is the beam width for all space, where $z_R = kw_0^2/2$ is the Rayleigh length, $R(z) = z\bigl[1 + (z_R/z)^2\bigr]$ is the radius of curvature, $\xi(z) = \arctan(z/z_R)$ is the Gouy phase, and $\mathcal{L}^{\ell}_p$ is the generalized Laguerre polynomial of azimuthal order $\ell$ and radial order $p$. With Eq.~(\ref{eq: acoustic-fields}), we derive the Cartesian velocity field components
\begin{align}{\label{eq: LG-velocity}}
    V_x(\mathbf{r}) &= -\frac{iP^{\text{(LG)}}}{\varrho_0 \omega}\Biggl(\frac{|\ell|}{\rho}e^{-i\varphi} + \rho\cos\varphi\Biggl[\frac{k}{R(z)} - \frac{2}{w^2(z)}\Biggr]\Biggr),\\
    V_y(\mathbf{r}) &= -\frac{iP^{\text{(LG)}}}{\varrho_0 \omega}\Biggl(\frac{|\ell|}{\rho}e^{i\varphi} + \rho\sin\varphi\Biggl[\frac{k}{R(z)} - \frac{2}{w^2(z)} \Biggr]\Biggr), \\
    V_z(\mathbf{r}) &= \frac{P^{\text{(LG)}}}{\varrho_0\omega}\Biggl\{\frac{iw_0^2}{z_R^2 w^2(z)}\Biggl[z+|\ell|z - \frac{2\rho^2z}{w^2(z)}\Biggr] \\
    &\quad+k\Biggl[1+\frac{\rho^2z_R^2}{z^2R^2(z)} - \frac{\rho^2}{2zR(z)} - \frac{w_0^2(|\ell|+2p+1)}{kz_R w^2(z)}\Biggr]\Biggl\}. \nonumber
\end{align}

Observe that the longitudinal field component $V_z$ vanishes along the propagation axis, i.e., where $(x,y)=(0,0),$ leaving nonzero transverse polarization that persists for infinite distance. This phenomena is illustrated in Fig.~\ref{fig: LG-plots}(a), where energy density $W$ is nonzero at the phase singularity beyond the Rayleigh length. The significance of nonzero intensity at the center of a vortex beam is widely studied in optics \cite{forbes2021}, particularly in light-matter interactions \cite{quinteiro2017, afanasev2020}. %As shown in the amplitudes at $z=0$ and $z=2zR$ in Fig.~\ref{fig: LG-plots}(a), the transverse field magnitudes shrink in theas the energy density $W$ diverges in space.%
 It should be noted that this effect is not found in acoustic Gaussian beams ($\ell=0$), which do not possess field nodes.

A straightforward analytical calculation (Eq.~(\ref{eq: norm-spin-density})) shows that for $\ell\neq 0$, the normalized spin AM density at the center of the vortex beam is strictly longitudinal $\mathbf{S}=S_z\mathbf{\hat{z}}.$ As shown in Fig.~\ref{fig: LG-plots}(b), $S_z=\pm1$ at the vortex center, corresponding to a line of circular polarization, of which chirality is determined by $\text{sgn}(\ell)$. It is found that $S_z$ persists for infinite distance, as presented in  Fig.~\ref{fig: LG-plots}(b), where $|\mathbf{S}|$ plateaus at unity in the vortex center at $z=0$ and $z=2z_R$. Due to the circular symmetry of the Laguerre-Gaussian beam, only a single line of circular polarization is present along the vortex node. Equivalent results for spin AM density can be derived for acoustic Bessel beams \cite{bliokh2019}, which possess propagation-invariant energy profiles.

Examining the LT-alignment parameter $\chi_{\text{LT}}$ near the phase singularity, $\rho\ll w(z)$, we find that 
\begin{equation}{\label{eq: chi-LG}}
\chi_{\text{LT}} \simeq \frac{2\xi^2 - 1}{2\xi^2+1},
\end{equation}
where $\xi = \ell/(k\rho).$ As shown in Fig.~\ref{fig: LG-plots}(c), the bounds for $\chi_{\text{LT}}$ are $-1 \leq \chi_{\text{LT}} \leq 1.$ The contour $\chi_{\text{LT}} = 0$ for any acoustic vortex beam of azimuthal order $\ell$ traces out a cylinder of fixed radius $\ell\lambda/(\sqrt{2}\pi)$ that encloses the vortex phase singularity for infinite distance, despite an expanding transverse intensity profile (see Fig.~\ref{fig: LG-plots}(a)).

 These non-diffractive features survive in the far-field for velocity fields of local coaxial and non-coaxial superpositions of monochromatic acoustic Laguerre-Gaussian beams, that is, $\mathbf{V}(\mathbf{r}) = \sum_i \mathbf{V}_{\ell_i, p_i}(\mathbf{r}_i).$ Indeed, when $z\gg D,$ where $D$ is the size of the source, the superimposed field components near a persisting longitudinal field zero gives quadratic behavior akin to Eq.~(\ref{eq: chi-LG}). Therefore, regardless of the localized source geometry and mixing of radial and azimuthal orders of acoustic vortex beams, there always exists non-diffractive features in distances sufficiently far from the source. We illustrate this phenomenon in Figures~\ref{fig: LG-plots}(d)-(f), which contain intensity and polarization distributions for the non-coaxial superposition of two acoustic vortex beams launched at $(x ,y)=(\pm 1.5\lambda, 0)$. 
 Around the longitudinal node propagating at $(x,y)=(0,0)$, there exists a pair of C lines that maintain separation of $\lambda/\pi$ (Fig.~\ref{fig: LG-plots}(e)). Furthermore, as evident in Fig.~\ref{fig: LG-plots}(f), interference of the individual fields result in a far-field non-diverging LT-alignment tube centered around the origin. 

\section{\label{sec: dipole-array}Acoustic Dipole Array}

Consider the pressure field of a dipole with moment $\mathbf{D}$, which can be expressed in the form \cite{wei2020}
\begin{equation}\label{eq: dipole-pressure}
P^{\text{dip}}(\mathbf{r}) = -\frac{i}{k}\mathbf{D} \cdot \nabla \Bigl(\frac{e^{ikr}}{kr}\Bigr).
\end{equation}
Note that the zeros of Eq.~(\ref{eq: dipole-pressure}) occur in the plane orthogonal to $\mathbf{D}.$ Thus, for a single dipole, there exists degenerate spin structures that enclose the plane where $P = 0.$ For example, when $\mathbf{D} = D\mathbf{\hat{z}},$ a Taylor expansion around $\theta=\pi/2$ of the velocity components $\mathbf{V}^{\text{dip}} = (V_r, V_{\theta}, V_{\varphi})^{\text{T}},$ which can be derived from Eq.~(\ref{eq: acoustic-fields}), gives the non-divergent contour $kr(\theta - \pi/2) = 1,$ independent of azimuthal coordinate $\varphi.$ Note that $r_{\parallel} = r(\theta - \pi/2)$ represents the longitudinal distance above the $z=0$ plane, so separation of the enclosing contour planes $\chi_{\text{LT}} = 0$ in this case is $2r_{\parallel} = \lambda /\pi.$ 

\begin{figure*}
    \includegraphics[width=7in]{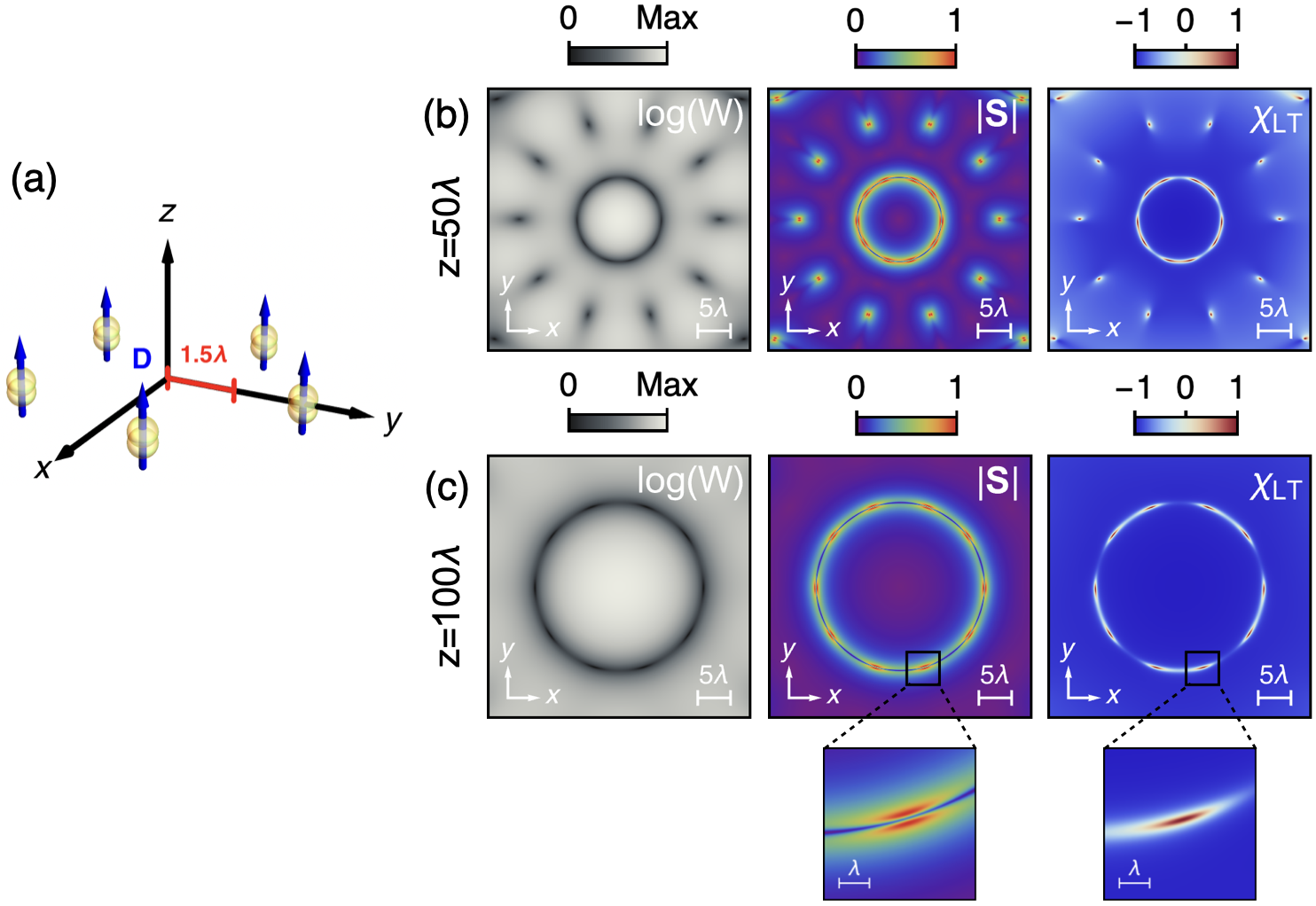}
    \caption{(a) Schematic diagram of wavelength-spaced dipoles, each with moment $\mathbf{D} = D\mathbf{\hat{z}},$ arranged in a pentagonal geometry centered around $(x,y)=(0,0)$ in the $z=0$ plane. (b) Density plots of logarithmic energy density $\log(W)$, normalized spin AM density $|\mathbf{S}|,$ and LT-alignment parameter $\chi_{\text{LT}}$ for the superimposed fields directly above the dipole array at $z=50\lambda.$ In (c), the same quantities are shown at $z=100\lambda.$ Inserts for $|\mathbf{S}|$ and $\chi_{\text{LT}}$ are shown around a propagating zero at $z=100\lambda.$}
    \label{fig: dipole-array-plots}
\end{figure*}

Here, we examine the superimposed acoustic fields from an array of acoustic dipoles, that is, $P = \sum_i P^{\text{dip}}(\mathbf{r}_i)$ and $\mathbf{V} = \sum_i \mathbf{V}^{\text{dip}}(\mathbf{r}_i),$ where $\mathbf{r}_i$ represents the spatial position of the $i$-th dipole. In the far field, the angular array factor $A(\theta, \varphi)$ can be decomposed from $P$ and $\mathbf{V}$, which gives the angular positions $(\theta_0, \varphi_0)$ of persisting field zeros. It follows that the array geometry, described by $A(\theta, \varphi)$, determines angular placement of the non-diffracting polarization and spin structures outlined in the formalism in Sec.~\ref{sec: formalism}.

As an example, we consider a pentagonal array of $\mathbf{\hat{z}}$-directed acoustic dipoles distributed over a circle of radius $3\lambda$ in the $z=0$ plane (see Fig.~\ref{fig: dipole-array-plots}(a) for a schematic diagram). Importantly, we do not neglect the transverse components $\mathbf{V}_{\perp} = (V_{\theta}, V_{\varphi})^{\text{T}}$ of $\mathbf{V}$, for their relative size is appreciable near the zeros of $V_r,$ which enables a nonparaxial description of the velocity fields at certain angular positions in space. As evident in Figs.~\ref{fig: dipole-array-plots}(b, c), the transverse energy density $W$ spreads in space with increasing distance $z$ from the array. Around each propagating dark spot of intensity, we have pairs of C-lines which maintain invariant separation of $\lambda/\pi$ with respect to radial distance. Moreover, enclosing each field node is the non-divergent elliptical tube expressed by the relation $|\mathbf{V}_{\perp}| =  |V_r|.$

\section{\label{sec: double-slit}Double-Slit Interference}

Our last example is the acoustic version of famous Young's Double Slit experiment of light interference, which was performed in 1801 and subsequently opened a new avenue for the study of wave optics. We provide a standard theoretical analysis of the experiment, with the inclusion of the 3D velocity field components typically neglected due to the curl-free behavior of sound. It follows that non-divergent pairs of circular polarization singularities sandwich the intensity minima (or dark fringes)  of the double-slit interference pattern.

\begin{figure}
    \includegraphics[width=3.4in]{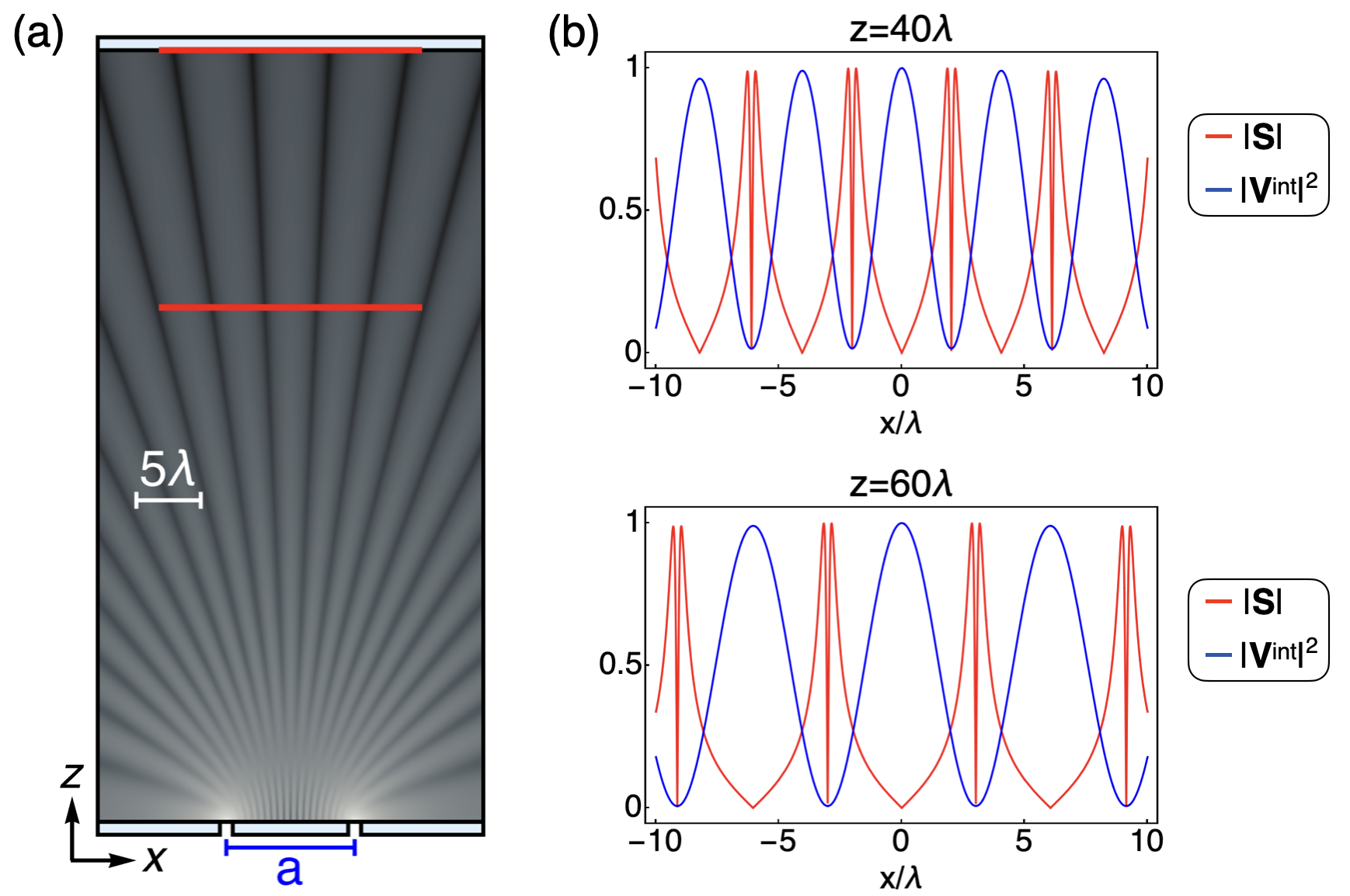}
    \caption{(a) Sketch of the double-slit interference geometry. The gray-scale distribution represents the logarithmic intensity of $|\mathbf{V}^{\text{int}}|$. (b) Amplitude plots of spin AM density $|\mathbf{S}|$ (red) and normalized intensity $|\mathbf{V}^{\text{int}}|^2$ (blue) at $z=40\lambda$ and $z=60\lambda,$ whose spatial positions with respect to the slits are indicated by the red lines in (a). Pairs of circular polarization singularities or C-lines ($|\mathbf{S}|=1$ corresponding to $S_y=\pm 1$) separated by $\lambda/\pi$ along $\mathbf{\hat{x}}$-direction sandwich each intensity minima at angle $\alpha_{\text{min}}$ in the far-field zone. Angular positions of C-lines are off-set by $\delta \alpha=\pm 1/zk$ with respect to $\alpha_{\text{min}}$.}
    \label{fig: double-slit-plots} 
\end{figure}

As depicted in Fig.~\ref{fig: double-slit-plots}(a), the experimental geometry is as follows. Consider a screen that consists of two narrow slits, both oriented in the $\mathbf{\hat{y}}$-direction, with spacing $a$ in the $\mathbf{\hat{x}}$-direction. A projection screen (parallel to $xy$-plane) is placed a distance $z$ from the slits. In the $y=0$ plane, the velocity field at a position $x$ on the projection screen can be expressed as the superposition of monochromatic point sources from each slit:
\begin{equation}{\label{eq: int-pressure}}
\mathbf{V}^{\text{int}}(\mathbf{r}) = \mathbf{V}_{10} \frac{e^{ikr_1}}{r_1} + \mathbf{V}_{20} \frac{e^{ikr_2}}{r_2},
\end{equation}
where $r_{1,2}$ represents the distance between each slit and the screen measurement position, and $\mathbf{V}_{10,20} = V_0(\sin \theta_{1,2}, 0, \cos \theta_{1,2})^{\text{T}} \simeq V_0(\theta_{1,2}, 0, 1)^{\text{T}}$ in the Cartesian basis, with the condition $z\gg x,a.$ The path difference $r_2 - r_1 = 2\pi m/k$, where $m$ is an integer, indicates constructive interference and a bright fringe; likewise, destructive interference occurs at $r_2 - r_1 = \pi (2m+1)/k$, where the dark fringe forms. It follows that the intensity maxima and minima, respectively, occur at the angles $\alpha_{\text{max}} \equiv x_{\text{max}}/z = \lambda m/a$ and $\alpha_{\text{min}} \equiv x_{\text{min}}/z = \lambda(2m+1)/(2a).$ For large distances from the slit screen, the longitudinal velocity field component $V_z^{\text{int}}$ is dominant in comparison to the transverse $V_x^{\text{int}}$, however, evaluating $\mathbf{V}^{\text{int}}$ near $\alpha_{\text{min}}$ results in comparable spatial field components. Noting that the angular position of each slit occurs at $\alpha_{1,2} = (x \mp a/2)/z,$  the ratio of velocity field components is derived to be
\begin{equation}{\label{eq: LT-int}}
\frac{V_x^{\text{int}}}{V_z^{\text{int}}} = \frac{x}{z} + \frac{a}{2z}\frac{1 - e^{ik(r_1-r_2)}}{1 + e^{ik(r_1-r_2)}}.
\end{equation}
In the vicinity of $x_{\text{min}},$ i.e., locations where $V_z^{\text{int}} = 0,$ the Taylor expansion $e^{ik(r_1-r_2)} \simeq -1 + ik\delta r$ results in $V_x^{\text{int}}/V_z^{\text{int}} \to \pm i\infty$ as $\delta r = r_1 - r_2 \to \mp 0,$ so $V_z^{\text{int}}$ is a first-order zero at $x_{\text{min}}.$ Moreover, the complex ratio $V_x^{\text{int}}/V_z^{\text{int}} = \mp i$, which satisfies the full circular polarization condition  $S_y=\pm 1$, occurs when $x_{\text{min}}$ is offset by $\delta x = \pm1/k=\pm \lambda/2\pi,$ {\it independently} of the distance between the slits' plane and the projection screen, corresponding to the path difference $\delta r = \pm a/(zk).$ Angular offset of the lines with $S_y=\pm 1$ from the dark fringe position is $\delta \alpha=\pm 1/zk$. That is, in the far-field interference pattern emanating from the double slits, we have pairs of circular polarization singularities (C-lines) of opposite spins that maintain a propagation-invariant separation  of $\lambda/\pi$ along $x$-axis across each propagating dark fringe of intensity, as demonstrated in Fig.~\ref{fig: double-slit-plots}(b).

 We notice an important difference from electromagnetism \cite{vernon2023}: While for electromagnetic double-slit interference the spinning structures do not form for field polarization aligned with the slits, there is no such a restriction for the longitudinal acoustic pressure waves.

\section{\label{sec: Conclusions}Summary and Conclusions}
We have presented a theoretical formalism that predicts non-diffractive polarization and spin structures inherent to acoustic fields of arbitrary sources. The far-field patterns of acoustic radiation yield dominant longitudinal velocity fields due to the curl-free nature of sound, however, near longitudinal zeros (i.e., pressure phase singularities) there exists a rich and invariant nonparaxial region that extends to infinity. In general, accompanying each acoustic field node are: a non-divergent elliptical tube whose contour is defined by equality of the longitudinal and transverse velocity fields and pairs of circular polarization singularities with constant sub-wavelength separation. The developed formalism can be extended to elastic waves in solids that would also involve shear waves known to lead to polarization anomalies \cite{helbig1987anomalous, lee2017off}. The next step in our studies may be an analysis of surface acoustic waves (SAW), for which the  polarization singularities predicted here could result in rotation of the velocity field plane near dark fringes of interference.

It is worth comparing our results to those found in electromagnetic fields \cite{afanasev2023, vernon2023}. Despite the contrasting transverse behavior of electric $\mathbf{E}$ and magnetic $\mathbf{H}$ fields ($\nabla \cdot \mathbf{E} = \nabla \cdot \mathbf{H} = 0$), it was shown that their transverse field zeros generate spinning structures analogous to our present work in acoustics.

  Thus, our findings offer fruitful analogies between light and sound, in regards to the interplay of longitudinal and transverse fields near dark spots of intensity. Moreover, we exemplified our general theory with acoustic vortex beams, acoustic dipole arrays, and Young's double slit experiment. We are hopeful this work provides potential applications in polarization-based communications, acoustic enantioseparation, non-destructive evaluation of materials, and acoustic tweezers.

\begin{acknowledgments}
We wish to acknowledge support from US Army Research Office Award No. W911NF-23-1-0085. AA thanks Michael Berry and Konstantin Bliokh for useful discussions.
\end{acknowledgments}

\bibliography{bibliography}

\end{document}